\documentclass[11pt,sort&compress]{elsarticle}
\usepackage[paper=letterpaper,top=35mm, bottom=35mm, left=35mm, right=35mm]{geometry}
\makeatletter
\def\ps@pprintTitle{%
 \let\@oddhead\@empty
 \let\@evenhead\@empty
 \def\@oddfoot{\centerline{\thepage}}%
 \let\@evenfoot\@oddfoot}
\makeatother
\usepackage{amsmath}
\usepackage{accents}
\usepackage{amssymb}
\usepackage{mathrsfs}
\usepackage[LGR,T1]{fontenc}
\usepackage[latin1]{inputenc}
\let\oldbibliography\thebibliography
\renewcommand{\thebibliography}[1]{%
  \oldbibliography{#1}%
  \setlength{\itemsep}{2pt}%
}
\usepackage[cal=boondox,calscaled=1]{mathalfa}
\DeclareMathAlphabet{\bbvar}{U}{BOONDOX-ds}{m}{n}
\DeclareMathAlphabet{\bbgreek}{U}{bbold}{m}{n}
\usepackage[defaultsans,osfigures,scale=0.9]{opensans}
\usepackage{graphicx}
\usepackage{psfrag}
\usepackage{pstool}
\usepackage{caption}
\usepackage{graphicx}%
\newcommand{\hook}{\text{\large{$\lrcorner$}}}
\usepackage{color}
\usepackage{hyperref}
\newcommand{\qq}[1]{``#1''} 
\newcommand{\q}[1]{`#1'\,}  

\newcommand{\di}{\mathrm{d}}
\usepackage{tensor}\newcommand{\ou}[3]{{#1}{}^{#2}{}_{#3}}
\newcommand{\uo}[3]{{#1}{}_{#2}{}^{#3}}

\newcommand{\I}{\mathrm{i}} 
\newcommand{\E}{\mathrm{e}} 
\newcommand{\CC}{\mathrm{cc.}} 
\newcommand{\C}{\mathbb{C}}
\newcommand{\N}{\mathbb{N}}
\newcommand{\R}{\mathbb{R}}
\newcommand{\Z}{\mathbb{Z}}

\newcommand{\1}{\mathnormal{1}}
\newcommand{\0}{o}

\newenvironment{subalign}{\subequations\align}{\endalign\endsubequations}
\newcommand{\eref}[1]{(\ref{#1})}

\DeclareMathAlphabet{\sfit}{OT1}{fos}{sb}{it}
\DeclareMathAlphabet{\mathsf}{OT1}{fos}{sb}{n}

\definecolor{darkgreen}{rgb}{0.01, 0.75, 0.24}

\newcommand{\normord}[1]{\boldsymbol{:}\mathrel{#1}\boldsymbol{:}}
\usepackage[multiple, flushmargin]{footmisc}

\begin{document}

\begin{abstract}
In this paper, we will make an attempt to clarify the relation between three-dimensional euclidean loop quantum gravity with vanishing cosmological constant and quantum field theory in the continuum. We will argue, in particular, that in three spacetime dimensions the discrete spectra for the geometric boundary observables that we find in loop quantum gravity can be understood from the quantisation of a conformal boundary field theory in the continuum without ever introducing spin networks or triangulations of space. At a technical level, the starting point is the Hamiltonian formalism for general relativity in regions with boundaries at finite distance. At these finite boundaries, we choose specific conformal boundary conditions (the boundary is a minimal surface) that are derived from a boundary field theory for an $SU(2)$ boundary spinor, which is minimally coupled to the spin connection in the bulk. The resulting boundary equations of motion define a conformal field theory with vanishing central charge. We will quantise this boundary field theory and show that the length of a one-dimensional cross section of the boundary has a discrete spectrum. In addition, we will introduce a new class of coherent states, study the quasi-local observables that generate the quasi-local Virasoro algebra and discuss some strategies to evaluate the partition function of the theory.
\end{abstract}

\title{Conformal boundary conditions, loop gravity and the continuum}
\author{Wolfgang Wieland}
\address{Perimeter Institute for Theoretical Physics\\31 Caroline Street North\\ Waterloo, ON N2L\,2Y5, Canada\\{\vspace{0.5em}\normalfont August 2018}
}
\maketitle
{\vspace{-1.2em}

{\tableofcontents}
\begin{center}{\noindent\rule{\linewidth}{0.4pt}}\end{center}\newpage
\section{Introduction}
\noindent In loop quantum gravity, geometric observables such as areas and angles have discrete spectra \cite{ashtekar, rovelli, thiemann}. These results were developed originally in the spin network representation, where quantum three-geometries are created by successively exciting gravitational Wilson loops out of a vacuum that represents no space at all \cite{Ashtekar:1993wf,Ashtekar:1994mh}. The appearance of such discrete and combinatorial structures led to the idea that the whole approach is some sort of gravitational lattice gauge theory and that the discrete spectra are a mere lattice artefact that should disappear once we found the physical Hilbert space in the continuum, cf.\ \cite{outside,Dittrich:2007th}. In this article we will demonstrate that this concern is unfounded: we will show that the discrete spectra of geometric observables for euclidean quantum gravity in three dimensions can be recovered from the quantisation of a conformal boundary field theory in the continuum without ever introducing spin networks or triangulations of space.\footnote{A generalisation of our argument to four Lorentzian spacetime dimensions can be found in \cite{Wieland:2017cmf}.}\vspace{0.1em}  

In the following, we will only be concerned with the theory for a vanishing cosmological constant. In addition, the entire calculation happens in a quasi-local context, where the gravitational field is quantised in a region with boundaries at finite distance. From the perspective of general covariance, this quasi-local approach may seem, in fact, more appealing than an asymptotic quantisation, because the physical size of a region (or the distance from a source) must be itself the outcome of a measurement, and there is no reason a priori why we should restrict ourselves to only those measurement outcomes for which the observer would be sitting at infinity. The limit to infinity would then only be performed within the quantum theory by selecting, for example, an appropriate sequence of coherent boundary states. We will propose such coherent states in the paper.\vspace{0.1em}  


The paper itself is divided into two parts: The first part develops the classical bulk plus boundary field theory. In the interior of spacetime, the theory is topological,\footnote{In three dimensions, general relativity has as many gauge constraints as there are configuration variables.} and there are no physical degrees of freedom in the bulk. The presence of a boundary changes the situation quite significantly: at the boundary, otherwise unphysical gauge directions turn into actual physical boundary degrees of freedom \cite{Carlip:2005zn,carlipbook,Brown:1986nw}. One of the key results of this paper is that the resulting gravitational edge modes \cite{Wieland:2017cmf,Wieland:2017ksn,Wieland:2017zkf,Speranza:2017gxd,Geiller:2017xad} can be neatly characterised by an $SU(2)$ boundary spinor $\xi^A$ that is minimally coupled to the spin connection in the bulk. The $SU(2)$ norm $\|\xi\|^2=\xi^\dagger_A\xi^A$ of this boundary spinor turns out to have a very simple geometric interpretation: it determines the conformal factor that relates the fiducial two-dimensional boundary metric to the physical metric in the bulk. The boundary field equations, which define a conformal field theory with vanishing central charge, turn out to have a neat interpretation as well: they impose that the boundary is a minimal surface with respect to the three-dimensional physical metric in the bulk (hence they determine how the boundary is embedded into the bulk). The second half of the paper deals with some aspects of the resulting quantum theory. We will show, in particular, that the physical length of a cross section of the boundary turns into the number  operator that counts the number of quanta that constitute the quantum boundary geometry.


\section{Boundary field theory at finite distance}
\subsection{Action, boundary conditions and field equations}\label{sec2.1}
\noindent For three-dimensional euclidean gravity with vanishing cosmological constant, we can work with the action
\begin{equation}
S_{\mathcal{M}}[e,A]=\frac{1}{8\pi G}\int_{\mathcal{M}}e_i\wedge F^i[A],\label{bulkterm}
\end{equation}
where $G$ is Newton's constant in three spacetime dimensions,\footnote{In units of $\hbar=c=1$, Newton's constant has dimensions of length, and $\ell_o:=4\pi G$ is the minimal Planck length in three dimensions.} $F^i[A]=\di\wedge A^i+\frac{1}{2}\ou{\epsilon}{i}{jk}A^j\wedge A^k$ is the curvature of the $SU(2)$ spin connection and $e_i$ is the co-triad with corresponding euclidean spacetime metric $g_{ab}=\delta_{ij}\ou{e}{i}{a}\ou{e}{j}{b}$. On shell, this metric is locally flat, which is a consequence of the equations of motion in the bulk, namely the vanishing of curvature ($F^i=0$) and torsion ($T^i=\nabla\wedge e^i=\di\wedge e^i+\ou{\epsilon}{i}{jk}A^j\wedge e^k=0$).

Suppose then that $\mathcal{M}$ has the topology of an infinite solid cylinder with boundary $\mathcal{B}=\partial\mathcal{M}\simeq \R\times S^1$. The goal of the article is to treat the resulting bulk plus boundary theory as a Hamiltonian system, quantise it and compare the results with what we know from loop quantum gravity.

Since we now have a boundary (at finite distance), we have to specify the boundary conditions, otherwise the variational principle is incomplete. Different boundary conditions require then different boundary terms, which, in turn, lead to different boundary field theories. In our case, and to facilitate the comparison with recent developments in loop quantum gravity \cite{Wieland:2017ksn,Wieland:2017zkf}, we choose the following conformal boundary conditions,
\begin{subalign}
&\varphi^\ast_{\mathcal{B}}g_{ab}\in[q_{ab}]\Leftrightarrow\exists \Omega:\mathcal{B}\rightarrow \R_+:\varphi^\ast_{\mathcal{B}}g_{ab}=\Omega^{-2}q_{ab}\label{bndryb},\\
&K=\nabla_an^a=0.\label{bndrya}
\end{subalign}
The first condition \eref{bndryb} says that the pull-back $\varphi^\ast_{\mathcal{B}} g_{ab}$ of the physical metric to the boundary $\mathcal{B}$ lies in the conformal class of some given and fiducial two-dimensional background metric $q_{ab}$ on $\mathcal{B}$. The second condition says that the trace of the extrinsic curvature of the boundary vanishes ($n^a$ is the normal vector to the boundary and $\nabla_a$ is the covariant derivative in the bulk). This condition is the same as to say that the boundary is an minimal surface with respect to the metric in the interior. The conformal factor and the trace free part of the extrinsic curvature are unconstrained.

In the following, both the boundary conditions (\hyperref[bndrya]{2a,b}) and the field equations in the interior will be derived from the variation of a coupled bulk plus boundary action.  The coupled system evolves according to a Hamiltonian that generates boundary data compatible with (\hyperref[bndrya]{2a,b}) for some initial datum on a \emph{one-dimensional} cross section $\mathcal{C}$ of the boundary $\mathcal{B}$. The resulting theory represents a concrete and quasi-local realisation of the holographic principle \cite{Maldacena:1997re,Witten:1998qj,Ammon:2015wua} for vanishing cosmological constant and boundaries at finite distance. The possibility of such a quasi-local approach is supported by most recent results in non-perturbative quantum gravity: Dittrich, Livine and collaborators have studied very recently the Ponzano\,--\,Regge amplitudes for discretised manifolds with cylindrical boundaries \cite{Dittrich:2017hnl,Dittrich:2017rvb,Dittrich:2018xuk} at finite distance\footnote{The series \cite{Dittrich:2017hnl,Dittrich:2017rvb,Dittrich:2018xuk} by Dittrich, Livine and collaborators is based on the usual Dirichlet boundary conditions rather than the conformal boundary conditions (\hyperref[bndrya]{2a,b}).} and recovered within this setting the results from the perturbative approach \cite{Barnich:2015mui} for the partition function of euclidean quantum gravity in three dimensions.\vspace{0.5em}

To find an appropriate boundary action that can realise this principle in terms of spin connection variables, let us first make the following observation: \emph{There always exists a boundary spinor $\xi^A$ (a section of the spin bundle over the boundary) and a complex basis $(m_a,\bar{m}_a)$ of the complexified cotangent space $T^\ast \mathcal{B}_\C$ such that the pull-back of the triad to the boundary $\mathcal{B}$ assumes the following form,}
\begin{equation}
\varphi^\ast_{\mathcal{B}}\ou{e}{i}{a}=\frac{4\pi G}{\sqrt{2}}\, \xi^A\xi^B\uo{\sigma}{AB}{i}m_a+\CC,\label{glucond}
\end{equation}
where $\{\ou{\sigma}{A}{Bi}\}$ are the Pauli matrices,\footnote{$A,B,C,\dots=0,1$ are spinor indices, which are raised and lowered using the skew-symmetric epsilon spinors $\epsilon^{AB}$ and $\epsilon_{AB}$. Our conventions are the following: $\xi_A=\xi^B\epsilon_{BA}$ and $\xi^A=\epsilon^{AB}\xi_B$. In particular $\sigma_{ABi}=\sigma_{BAi}=\epsilon_{CB}\ou{\sigma}{C}{Ai}$, and $\xi_A\xi^A=0$.} and the factor of $4\pi G$ has been introduced for later convenience. The simplest way to see that such a pair $(\xi^A,m_a)$ can always be found is the following: first of all, we note that the pull-back of the metric to the boundary can always be written in terms of a conformal factor\footnote{We could always achieve $\Omega=1$, but for the whole purpose of this paper it is useful to be more general.}} $\Omega:\mathcal{B}\rightarrow\R_+$ and a complex dyad $(m_a,\bar{m}_a)$ as
\begin{equation}
\varphi^\ast_{\mathcal{B}}g_{ab}=\Omega^{-2} (m_{a}\bar{m}_{b}+\bar{m}_{a}m_{a})\equiv \Omega^{-2} q_{ab}.\label{Omdef}
\end{equation}
Denote then by $(m^a,\bar{m}^a)\in T\mathcal{B}_\C$ the dual basis, $m^a\bar{m}_a=1$, $m^am_a=0$. Next, construct the following element of the complexified Lie algebra $\mathfrak{su}(2)_\C$, namely
\begin{equation}
\ou{\bar m}{A}{B}=\frac{1}{\sqrt{2}}\ou{\sigma}{A}{Bi}\bar{m}^a\varphi^\ast_{\mathcal{B}}\ou{e}{i}{a}.
\end{equation}
Since $\bar{m}^a$ is a complex null vector with respect to the boundary metric $\varphi^\ast_{\mathcal{B}}g_{ab}$, we now also know that $\ou{\bar m}{A}{C}\ou{\bar m}{C}{B}=0$, which implies that there exists a spinor $\xi^A$, such that
\begin{equation}
\ou{\bar{m}}{A}{B}\propto\xi^A\xi_B.\label{barm}
\end{equation}
Comparison with \eref{glucond} completes the argument. Going back to \eref{Omdef}, we can then also see that the conformal factor must be proportional to the $SU(2)$ norm $\|\xi\|^2=\delta_{AA'}\bar{\xi}^{A'}\xi^A$ of the boundary spinor. A short calculation fixes the overall normalisation,
\begin{equation}
\Omega^{-1}=4\pi G\,\|\xi\|^2.
\end{equation}

To introduce an appropriate boundary action, we now only have the boundary dyad $(m_a,\bar{m}_a)$, the boundary spinor $\xi^A$, and its covariant exterior derivatives at our disposal. Going back to equation \eref{glucond}, we  consider, therefore, the following candidate for a bulk plus boundary action, namely
\begin{equation}
S[e,A|\xi]=\frac{1}{8\pi G}\int_{\mathcal{M}}e_i\wedge F^i[A]-\frac{\I}{\sqrt{2}}\int_{\mathcal{B}}\Big(\xi_Am\wedge D\xi^A-\CC\Big),\label{fullactn}
\end{equation}
where $D_a$ is the pull-back of the $SU(2)$ covariant derivative to the boundary,
\begin{equation}
D_a\xi^A=\partial_a\xi^A+\frac{1}{2\I}\ou{\sigma}{A}{Bi}(\varphi^\ast_{\mathcal{B}}\ou{A}{i}{a})\xi^B.
\end{equation}
In the following, we will treat the coupled bulk plus boundary geometry as one dynamical system, whose histories are given by the triple $(\ou{e}{i}{a},\ou{A}{i}{a},\xi^A)$ of bulk plus boundary fields. The complex boundary dyad $(m_a,\bar{m}_a)$, on the other hand, is treated as an external background structure, whose field variations vanish,
\begin{equation}
\delta m_a=0.\label{backgrnd}
\end{equation}

Having introduced additional boundary fields, we now get additional boundary equations of motion (in addition to the flatness constraint and the vanishing of torsion in the bulk). First of all, there are the \emph{glueing conditions} linking the fields in the interior to the spinors at the boundary. Using $\delta F=\nabla\wedge\delta A$ together with Stokes's theorem, we calculate the variation of the action \eref{fullactn} with respect to the spin connection, obtaining
\begin{align}\nonumber
\delta_A S[e,A|\xi]=\frac{1}{8\pi G}&\int_{\mathcal{M}}(\nabla \wedge e_i)\wedge\delta A^i+\\
&-\frac{1}{8\pi G}\int_{\mathcal{B}}\Big[e_i-\frac{4\pi G}{\sqrt{2}}\Big(m\xi^A\xi^B\sigma_{ABi}+\CC\Big)\Big]\wedge \delta A^i.\label{convar}
\end{align}
Imposing $\delta_A S=0$, gives us therefore \emph{two} conditions, namely the torsionless condition ($\nabla\wedge e^i=0$) in the bulk and the glueing conditions \eref{glucond}, which are just the first part of our boundary conditions, namely \eref{bndryb}. The second part \eref{bndrya} of the boundary conditions follows from the variations of the boundary spinor $\xi^A$. The resulting boundary equations of motion require that the boundary spinor $\xi^A$ is holomorphic, namely
\begin{equation}
m^a\mathcal{D}_a\xi^A=0,\label{EOMb}
\end{equation}
where we introduced the $SU(2)\times U(1)$ gauge covariant derivative,
\begin{equation}
\mathcal{D}_a\psi^A:=D_a\psi^A+\frac{1}{2\I}\Gamma_a\psi^A,\label{coderiv}
\end{equation}
for the $U(1)$ spin connection $\Gamma_a\in T^\ast\mathcal{B}$ at the boundary. This connection is implicitly defined via the two-dimensional (boundary intrinsic) torsionless condition
\begin{equation}
\mathcal{D}_{[a}m_{b]}=\partial_{[a}m_{b]}+\I\Gamma_{[a}m_{b]}=0,
\end{equation}
and it can be extended naturally to tensor fields $\ou{T}{a\dots}{b\dots}$ intrinsic to the boundary: the resulting torsionless derivative $\mathcal{D}_a$ annihilates $q_{ab}$, as well as the $U(1)$ soldering forms $m_a$ and $\bar{m}_a$, e.g.\ $\mathcal{D}_am_b=0$. In the next section, we will show that the boundary field equations \eref{EOMb} are equivalent to the second part of the boundary conditions, namely \eref{bndrya}.
\subsection{The boundary as a minimal surface}\label{sec2.2}
\noindent The boundary equations of motion \eref{EOMb} have a simple geometric interpretation: they impose that the boundary $\mathcal{B}$ is a minimal surface with respect to the geometry in the interior, thus the extrinsic curvature vanishes: $K=\nabla_an^a=0$, such that the boundary conditions (\hyperref[bndryb]{2a,b}) both follow from the equations of motion of the bulk plus boundary theory.

To see that the boundary equations of motion \eref{EOMb} imply that the extrinsic curvature vanishes, consider first the following normalised internal three-vector, which is the \emph{\q{expectation value}} of the Pauli matrices with respect to the boundary spinor $\xi^A$, namely,
\begin{equation}
n_i=\frac{\xi^\dagger_A\ou{\sigma}{A}{Bi}\xi^B}{\|\xi\|^2},\label{ndef}
\end{equation}
where $\xi^\dagger_A=\delta_{AA'}\bar{\xi}^{A'}$ is the conjugate spinor with respect to the $SU(2)$ invariant hermitian metric. Going back to the glueing condition \eref{glucond} and recognising $\xi_A\xi^A=\epsilon_{BA}\xi^B\xi^A=0$, we see that $n_a=n^ie_{ia}$ is indeed the co-vector normal of the boundary (hence $\varphi^\ast_{\mathcal{B}}e^in_i=0$). We can then define immediately twist $\omega$ and expansion $K=\vartheta$ of $n_a$, namely
\begin{equation}
\vartheta:=2m^{(a}\bar{m}^{b)}\nabla_an_b,\quad \omega:=2\I m^{[a}\bar{m}^{b]}\nabla_an_b.
\end{equation}
Using the isomorphism between vectors and spinors, we find
\begin{equation}
\frac{\vartheta+\I\omega}{\sqrt{2}}=\sqrt{2}\,\bar{m}^bm^aD_an_b=8\pi G\,\xi^A\xi^Bm^aD_a\Big(\frac{\xi^\dagger_{A}\xi^{\phantom{\dagger}}_{B}}{\|\xi\|^2}\Big)={8\pi G}\,\xi^Am^aD_a\xi_A.\label{excurvatr}
\end{equation}
Now, the twist $\omega$ must vanish since $n^a$ is normal to a two-dimensional boundary of a three-dimensional manifold $\mathcal{M}$. For a minimal surface, $\vartheta$ (which is the trace of the extrinsic curvature) must vanish as well, which implies $m^aD_a\xi^A=f\xi^A$ for some function $f$. The function $f$ is further constrained by the pull-back of the torsionless condition $\nabla\wedge e^i=0$ to the boundary $\mathcal{B}$, which implies $f=\frac{\I}{2}m^a\Gamma_a$. The vanishing of the extrinsic curvature is therefore equivalent to the holomorphicity of the boundary spinor: $m^a\mathcal{D}_a\xi^A=0\Leftrightarrow K=0$, and these are just our boundary equations of motion that we found in above, see \eref{EOMb}. The boundary field equations \eref{EOMb} impose, therefore, that the boundary $\mathcal{B}=\partial\mathcal{M}$ is a minimal surface, hence $K=0$ as required by our boundary conditions \eref{bndrya}. Both boundary conditions (\hyperref[bndryb]{2a,b}) follow therefore from the variation of the bulk plus boundary action. 

\subsection{A minimal example: the boundary spinor of a catenoid}\label{sec2.3}
\noindent To gain some more intuition for the boundary field theory, we now want to find an explicit solution of the field equations. In the last section, we saw that the boundary of the cylinder $\mathcal{M}$ must be a minimal surface. In addition, we also know that the metric in the interior of the cylinder is flat. The simplest geometry that satisfies these conditions is a \emph{catenoid} flatly embedded in $\R^3$. Using coordinates $\eta\in\R$ and $\varphi\in[0,2\pi)$, we can parametrise this surface as follows,
\begin{align}\nonumber
x^1&=\rho\cosh\eta\cos\varphi,\\\nonumber
x^2&=\rho\cosh\eta\sin\varphi,\\
x^3&=\rho\,\eta,\label{catdef}
\end{align}
where $\rho> 0$ is a free parameter that determines the minimal radius of the catenoid, which will turn into an operator at the quantum level. 

To find the boundary spinor that satisfies the field equations \eref{EOMb} for this geometry, we proceed as follows.  On the surface of the catenoid, we introduce the complex coordinate
\begin{equation}
z=\E^{\eta+\I\varphi}.
\end{equation}
We now define the following one-forms and tangent vectors intrinsic to the tangent bundle of the boundary, namely,
\begin{equation}
m_a=\frac{1}{\sqrt{2}}\partial_az,\qquad \bar{m}^a=\sqrt{2}\,\partial^a_z=\frac{1}{\sqrt{2}}\E^{\eta+\I\varphi}\big(\partial^a_\eta-\I\partial^a_\varphi\big).
\end{equation}
Using cartesian coordinates $(x^1,x^2,x^3)$, we can represent the tangent vector $\partial^a_z$ as a column vector,
\begin{equation}
\partial_z\vec{x}=\frac{\rho}{2}\E^{-\eta-\I\varphi}\begin{pmatrix}\phantom{-\I\,}\sinh(\eta+\I\varphi)\\-\I\,\cosh(\eta+\I\varphi)\\1\end{pmatrix}.
\end{equation}
We now contract this vector with the Pauli matrices $\vec{\sigma}$. To solve the glueing conditions \eref{glucond}, we have to find a spinor $\xi^A$ such that,
\begin{equation}
\vec{\sigma}\cdot\partial_z\vec{x}=\frac{\rho}{2z}\begin{pmatrix}1 & -z^{-1}\\z&-1\end{pmatrix}=-4\pi G\begin{pmatrix}\xi^0\\\xi^1\end{pmatrix}\otimes\begin{pmatrix}-\xi^1&\xi^0\end{pmatrix}.
\end{equation}
This equation has two solutions for $\xi^A$, namely
\begin{equation}
\begin{pmatrix}\xi^0(z)\\\xi^1(z)\end{pmatrix}=\pm\sqrt{\frac{\rho}{8\pi G}}\begin{pmatrix}z^{-1}\\1\end{pmatrix},\label{xisolved}
\end{equation}
which is a holomorphic function of the boundary spinor, as it is indeed required by the boundary equations of motion \eref{EOMb}.

\subsection{Covariant Hamiltonian analysis, gauge symmetries}\label{sec2.4}
\noindent Our next task is to explore the solution space of the theory and study the gauge symmetries at the Hamiltonian level. First of all, we have to identify the symplectic structure on an \emph{\q{initial}}  hypersurface. We thus slice the cylinder $\mathcal{M}$ along a hypersurface $\Sigma$ into two halves. The first variation of the action (in either half) determines then the covariant pre-symplectic potential \cite{Wald:1999wa}, namely
\begin{equation}
\Theta_\Sigma=-\frac{1}{8\pi G}\int_\Sigma e_i\wedge\bbvar{d}A^i-\frac{\I}{\sqrt{2}}\oint_{\mathcal{C}}\big(\xi_A m\,\bbvar{d}\xi^A-\CC\big),\label{thetadef}
\end{equation}
where the symbol \qq{$\bbvar{d}$} is the exterior functional derivative on the classical solution space of the theory with\footnote{The orientation of $\mathcal{C}$ is induced from the orientation on $\Sigma$. The choice of orientation is a frequent source of sign mistakes.} $\mathcal{C}=\partial\Sigma$ denoting a one-dimensional cross section of the boundary $\mathcal{B}=\partial\mathcal{M}$. To study the symmetries of the theory at the Hamiltonian level, we introduce the pre-symplectic two-form,
\begin{equation}
\Omega_\Sigma=\bbvar{d}\Theta_\Sigma.
\end{equation}

Internal $SU(2)$ gauge transformations (generated by an $SU(2)$ gauge element $\Lambda^i$) act now in the obvious way on the bulk and boundary variables,
\begin{equation}
\delta_\Lambda e^i=\ou{\epsilon}{i}{lm}\Lambda^le^m,\quad\delta_\Lambda A^i=-\nabla\Lambda^i,\quad\delta_\Lambda\xi^A=\frac{1}{2\I}\ou{\sigma}{A}{Bi}\Lambda^i\xi^B,
\end{equation}
where $\nabla_a$ is the $SU(2)$ exterior covariant derivative in the bulk. A short calculation reveals that the vector field $\delta_\Lambda$ (on phase space) is indeed a degenerate direction of the pre-symplectic two-form,
\begin{equation}
\Omega_\Sigma(\delta_\Lambda,\cdot)=0,\label{Lgaugetrafo}
\end{equation}
hence $\delta_\Lambda$ generates gauge transformations at the Hamiltonian level. It is worth noticing that this is true even for those $SU(2)$ gauge transformations that do not vanish at the boundary ($\Lambda^i\big|_{\mathcal{B}}\neq 0$), and this observations will be important below. The explicit proof is straight forward, and can be found in an earlier paper of this series, see \cite{Wieland:2017ksn}.

\vspace{0.5em}
The situation for the boundary diffeomorphisms is more interesting. Consider first a vector field $t^a\in T\mathcal{M}$ that preserves the boundary,
\begin{equation}
t^a\big|_{\partial\mathcal{M}}\in T\mathcal{B}.
\end{equation}
Such a diffeomorphism can now act naturally on all our configuration variables. It acts on the variables in the bulk through the gauge covariant Lie derivative\footnote{The symbol \qq{$\hook$} denotes the interior product, $(t\hook\omega)_{bc\dots}=t^a\omega_{abc\dots}$} $\mathcal{L}_t$,
\begin{subequations}\begin{align}
\delta_t e^i&=\mathcal{L}_te^i=t\hook(\nabla\wedge e^i)+\nabla(t\hook e^i).\\
\delta_t A^i&=\mathcal{L}_tA^i=t\hook F^i,
\label{gaugedlie}\end{align}\end{subequations}
but the crucial question is how the vector field $\delta_t$ (on the covariant phase space) should act now on the boundary spinor $\xi^A$. The naive definition $\delta_t\xi^A:=\mathcal{L}_t\xi^A=t^a\mathcal{D}_a\xi^A$ (for the $SU(2)\times U(1)$ covariant derivative $\mathcal{D}_a$) gives an undesirable result. The reason is that the complex co-dyad $(m_a,\bar{m}_a)$ of $T^\ast \mathcal{B}$ is treated as an external background structure (a $c$-number in the old terminology), whose field variations vanish (see \eref{backgrnd}), but $\mathcal{L}_tm_a\neq 0$ for generic vector fields $t^a\in T\mathcal{B}$. The idea and solution of this trouble is to shuffle the burden from $m_a$ into the boundary variable $\xi^A$ (which is a $q$-number) by introducing a suitable correction term $\Delta_t\xi^A$. To find this missing term and define $\delta_t\xi^A$ as $\delta_t\xi^A:=t^a\mathcal{D}_a\xi^A+\Delta_t\xi^A$, consider first the pull-back to the boundary of the triad \eref{glucond}. We take the Lie derivative on both sides of the glueing conditions \eref{glucond}, and demand\footnote{The condition simply says that the field variation on the boundary phase space equals the Lie derivative on space time.} $\delta_t\varphi^\ast_{\mathcal{B}} e^i=\varphi^\ast_{\mathcal{B}} \mathcal{L}_te^i$. The gauged Lie derivative $\mathcal{L}_t$ annihilates the Pauli matrices. The field variation $\delta_t$, on the other hand, annihilates $m_a$. This implies the following condition for $\delta_t\xi^A$, namely
\begin{equation}
2(\delta_{t}\xi^{(A})\xi^{B)}\uo{\sigma}{AB}{i}m_a+\CC=2(\mathcal{L}_t\xi^{(A})\xi^{B)}\uo{\sigma}{AB}{i}m_a+\xi^A\xi^B\uo{\sigma}{AB}{i}\mathcal{L}_tm_a+\CC\label{cond}
\end{equation}
Using that $\xi^A$ and $\epsilon^{AB}\xi^\dagger_B=\ou{\delta}{A}{A'}\bar{\xi}^{A'}$ are linearly independent, we compare coefficients and conclude that
\begin{equation}
\Delta_t\xi^A\propto\xi^A.\label{Deltadef}
\end{equation}
and $\mathcal{L}_tm_a=\mathcal{D}_at^bm_b\propto m_a$. The latter is the same  as to say that $t^a$ must be a conformal Killing vector field at the boundary (i.e.\ $\mathcal{D}_{(a}t_{b)}\propto q_{ab}$). Going back to \eref{cond}, we can then fix the proportionality constant in \eref{Deltadef}, obtaining
\begin{equation}
\Delta_t\xi^A=\frac{1}{2}(\bar{m}^{a}\mathcal{L}_tm_a)\xi^A.
\end{equation}
Given any vector field $t^a\in T\mathcal{M}$, whose restriction to the boundary defines a conformal Killing vector field of the boundary metric $q_{ab}=2m_{(a}\bar{m}_{b)}$, we define, therefore, the field variation as follows,
\begin{equation}
\delta_t\xi^A:=t^a\mathcal{D}_a\xi^A+\frac{1}{2}(\bar{m}^{a}\mathcal{L}_tm_a)\xi^A,\label{confmap}
\end{equation}
which shows that the spinor $\xi^A$ is a conformal field of weight $(1/2,0)$.\footnote{Using holomorphic coordinates ($m_a=\frac{1}{\sqrt{2}}\partial_a z$), we can immediately integrate this vector field $\delta_t$ on field space such that $\xi^A$ transforms under a conformal map $f:\mathcal{B}\rightarrow\mathcal{B}$ as: $(f^\ast\xi^A)(z)=(\partial_z f)^{\frac{1}{2}}(z)\xi^A(f(z))$, for $f(z)=\exp(t)[z]$. } Since the classical action \eref{fullactn} is invariant under such conformal maps, the same conclusion could have been already made at the Lagrangian level. 

Having now defined the vector field $\delta_t$ on field space, we now want to show that it is generated by a Hamiltonian $H_t[\mathcal{C}]$. We thus want to integrate the Hamilton equations
\begin{equation}
\Omega_\Sigma(\delta_t,\delta)=-\delta H_t[\mathcal{C}],\label{Hameq}
\end{equation}
for all tangent vectors $\delta$ to the covariant phase space (linearised solutions $\delta\equiv(\delta A^i,\delta e_i,\delta\xi^A)$ of the field equations). First of all, we have
\begin{align}
\Omega_\Sigma(\delta_t,\delta)=-\frac{1}{8\pi G}\int_\Sigma\Big[\delta_te_i\wedge\delta A^i-\delta e_i\wedge \delta_tA^i\Big]-\I\sqrt{2}\oint_{\mathcal{C}}\!\Big[m\,\delta_t\xi_A\delta\xi^A-\CC\Big].
\end{align}
The field equations imply $\delta_t e_i=\nabla(t\hook e_i)$, and $\delta_t A^i=t\hook F^i=0$, and $\nabla\wedge\delta A^i=\delta F^i=0$, hence the first term is a total derivative. Using Stokes's theorem, we are thus left with a line integral over the circumference of the disk. If we finally also insert the glueing conditions \eref{glucond}, we find
\begin{align}\nonumber
\Omega_\Sigma(\delta_t,\delta)=-\frac{\I}{\sqrt{2}}&\oint_{\mathcal{C}}\Big[\frac{1}{2\I}\xi^A\xi^B\sigma_{ABi}N-\CC\Big]\delta A^i+\\
&-\frac{\I}{\sqrt{2}}\oint_{\mathcal{C}}\Big[2m(\mathcal{L}_t\xi_A)\delta\xi^A+m(\bar{m}^b\mathcal{D}_bN)\xi_A\delta\xi^A-\CC\Big],
\end{align}
where $t^a\big|_{\mathcal{B}}=N\bar{m}^a+\bar{N}m^a$. The right hand side is a total derivative: we introduce the quasi-local Hamiltonian,
\begin{equation}
H_t[\mathcal{C}]=-\frac{\I}{\sqrt{2}}\oint_{\mathcal{C}}\Big[N\xi_A\mathcal{D}\xi^A-\CC\Big],\label{Hdef1}
\end{equation}
and using that $\mathcal{C}$ is closed, we will find that $H_t[\mathcal{C}]$ indeed integrates the Hamilton equations \eref{Hameq} for the field variation $\delta_t$. In addition, $H_t[\mathcal{C}]$ is conserved since $N$ is holomorphic, i.e.\ $m^a\mathcal{D}_aN=0$. Geometrically, the Hamiltonian is the integral over the shear $\sigma$ of $n_a$ ($n_a$ being the co-vector normal  \eref{ndef} to the boundary),
\begin{equation}
\bar\sigma=\bar{m}^a\bar{m}^b\nabla_an_b=-4\pi G\sqrt{2}\xi_A\bar{m}^aD_a\xi^A,
\end{equation}
hence
\begin{equation}
H_t[\mathcal{C}]=-\frac{1}{4\pi G}\oint_{\mathcal{C}}\mathrm{Im}\big[m\,N\bar\sigma\big].
\end{equation}
This integral can also be written in terms of the canonical Brown\,--\,York \cite{BrownYork,Szabados:2004vb} quasi-local energy momentum tensor
\begin{equation}
8\pi G\,T_{ab}=-m_am_b\bar{\sigma}+\CC=-K_{ab},
\end{equation}
where $K_{ab}$ is the extrinsic curvature tensor\footnote{$K_{ab}$ is traceless, because the boundary is a minimal surface.} of the boundary $\partial\mathcal{M}$. We then have
\begin{equation}
H_t[\mathcal{C}]=\oint_{\mathcal{C}}dv^at^bT_{ab},
\end{equation}
with $dv^a=-\I\bar{m}^a\,m+\CC$ denoting the oriented line element on $\mathcal{C}=\partial\Sigma$.

\subsection{Covariant phase space, Virasoro generators, vanishing central charge}\label{sec2.5}
\noindent To quantise the theory, it is helpful to work in an explicit coordinate representation on the manifold, as we did in our example of the catenoid above (see \hyperref[sec2.3]{section 2.3}). First of all, we can always go to a gauge where the dyad $(m_a,\bar{m}_a)$ on the cylinder is given as
\begin{equation}
m=\frac{1}{\sqrt{2}}\di z,
\end{equation}
for coordinates $z=x+\I y$ on the boundary, such that the metric $q_{ab}=2m_{(a}\bar{m}_{b)}$ defines the fiducial line element $ds^2=\di x^2+\di y^2$ on $\partial\mathcal{M}=\mathcal{B}\simeq\C^2-\{0\}$. In addition, we may also assume that the interior of $\mathcal{M}$ is simply connected, that there are no defects or black holes, such that, in other words, we can always find a gauge such that the connection vanishes on $\mathcal{M}$,
\begin{equation}
\ou{A}{i}{a}=0. \label{gaugefix}
\end{equation}
This is only a partial gauge fixing.\footnote{Notice that even those large gauge transformations that do not vanish at the boundary are degenerate directions of the pre-symplectic two-form, see \eref{Lgaugetrafo} and \cite{Wieland:2017ksn,Wieland:2017zkf}.\label{gaugefoot}} Residual gauge transformations are generated by constant gauge elements $\Lambda^i:\partial_a\Lambda^i=0$. 

The boundary equations of motion \eref{EOMb} tell us then that the boundary spinor must be a holomorphic function of $z$,
\begin{equation}
\partial_{\bar z}\xi^A=0.
\end{equation}
Working with spinors, we have two possible boundary conditions on the cylinder,
\begin{equation}
\xi^A(\E^{2\pi\I}z)=\E^{2\pi\I\vartheta_\pm}\xi^A(z),
\end{equation}
with $\vartheta_+=0$ for the Neveu\,--\,Schwarz boundary conditions and $\vartheta_-=\tfrac{1}{2}$ for Ramond boundary conditions corresponding to the two possible spin structures on the circle $\mathcal{C}$. In the following, we will restrict ourselves to Neveau\,--\,Schwarz boundary conditions (as for the catenoid), because they correspond to the unique spin structure that the boundary $\mathcal{C}$ inherits from the disk $\Sigma$. We now have a Laurent expansion,
\begin{equation}
\xi^A(z)=\frac{1}{\sqrt{2\pi}}\sum_{n=-\infty}^\infty \xi^A_nz^n.\label{xiseries}
\end{equation}
If we go back then to the covariant symplectic potential \eref{thetadef}, we find
\begin{equation}
\Theta_\Sigma=\frac{1}{2}\sum_{n\in\Z}\epsilon_{AB}\xi^A_n\bbvar{d}\xi^{B}_{-n-1}+\CC
\end{equation}
This implies the fundamental Poisson commutation relations,
\begin{subequations}\begin{align}
\big\{\xi^A_m,\xi^B_n\big\}&=\epsilon^{AB}\delta_{m+n+1},\label{PoissStrct1}\\
\big\{\bar\xi^{A'}_m,\bar\xi^{B'}_n\big\}&=\bar\epsilon^{A'B'}\delta_{m+n+1},\label{PoissStrct2}\\
\big\{\xi^A_m,\bar\xi^{A'}_n\big\}&=0.\label{PoissStrct3}\end{align}
\end{subequations}

A conformal Killing vector admits now a Laurent expansion $t^a=\sum_{n\in\Z}t_nz^{n+1}\partial^a_{z}+\mathrm{cc}$. The corresponding quasi-local Hamiltonian \eref{Hdef1} is a sum of the Virasoro generators $L_n$ and $\bar{L}_n$,
\begin{equation}
H_{t}[\mathcal{C}]=\sum_{n=-\infty}^\infty\big(t_nL_n+\CC\big),\label{Hdef3}
\end{equation}
where\footnote{To avoid cluttering of indices, we write $\xi_A^n:=\epsilon_{BA}\xi^B_n$.}
\begin{equation}
L_n=\frac{1}{4}\sum_{m=-\infty}^\infty(2m+n+1)\xi_{A}^{-m-n-1}\xi^A_{m}.\label{Lndef}
\end{equation}
Using the Poisson brackets (\hyperref[PoissStrct1]{48a--c}), one can easily check that the generators satisfy the classical Virasoro algebra with vanishing central charge,
\begin{equation}
\big\{L_n,L_{n'}\big\}=(n-n')L_{n+n'},\quad\big\{\bar{L}_n,\bar{L}_{n'}\big\}=(n-n')\bar{L}_{n+n'}.
\end{equation}
The charges \eref{Lndef} can be immediately evaluated for the family of catenoids introduced in \hyperref[sec2.3]{section 2.3}. All but the lowest Virasoro generator vanish,
\begin{equation}
L_n=\frac{\boldsymbol{L}[\mathcal{C}_0]}{16\pi G}\delta_{n0},\label{LHam}
\end{equation}
where $\boldsymbol{L}[\mathcal{C}_0]=2\pi \rho$ is the minimal circumference of the catenoid. This equation says that the quasi-local energy $H=L_0+\bar{L}_0$ that we associate to the catenoid can be identified with the circumference of its bottleneck (divided by $8\pi G$) while its angular momentum $J=-\I(L_0-\bar{L}_0)$ vanishes (the geometry is rotationally symmetric). A very similar equation can also be found in (3+1)-dimensions, where the area of a surface equates to the generator of boosts into the orthogonal direction, see \cite{Carlip:1993sa,bwsurf,FGPfirstlaw,Wieland:2017zkf,DeLorenzo:2017tgx}.

\section{Boundary CFT over the Ashtekar\,--\,Lewandowski vacuum}
\subsection{Quantisation of geometry}\label{sec3.1}
\noindent The next step ahead is to quantise the classical phase space and study the resulting quantum geometries. First of all we note that the boundary spinors $\xi^A_n$ are bosons. This may sound a little odd, but it can be inferred immediately from our initial parametrisation \eref{glucond} of the triad at the boundary. Going back to the glueing conditions \eref{glucond}, we see that $\xi_A\xi_B=\xi_B\xi_A$, for otherwise we would have $\xi_A\xi_B\propto\epsilon_{AB}$, which would imply a fully degenerate triad at the boundary (namely $\varphi^\ast_{\mathcal{B}} \ou{e}{i}{a}=0$, which is impossible for any regular geometry). The boundary spinor $\xi_A$ must be, therefore, a boson taking values in the $\C^2$ spin bundle over the boundary. The classical Poisson algebra (\hyperref[PoissStrct1]{48a--c}) implies then the Heisenberg commutation relations at the quantum level. In units of $\hbar=1$,
\begin{equation}
\big[\xi^A_m,\xi^B_n\big]=-\I\epsilon^{AB}\delta_{m+n+1},
\end{equation}
and equally for the complex conjugate variables. At this point, it is now useful to introduce the following canonical coordinates. 
\begin{align}
z^A_m&:=\xi^A_m,\qquad p^A_m:=\xi^A_{-m-1},\qquad m=0,1,2,\dots,
\end{align}
which satisfy\footnote{Spinor indices are raised and lowered with the skew symmetric epsilon tensor, e.g.\ $p_A^n=\epsilon_{BA}p^B_n$.}
\begin{equation}
\big[p_A^n,z^B_m\big]=-\I\delta_A^B\delta_{mn}.
\end{equation}

For each individual mode, we may now choose a position representation on the Hilbert space $\mathcal{H}_n=L^2(\C^2,d^4z_n)$, where $z^A_n$ and $\bar{z}^{A'}_n$ act by multiplication while the conjugate moments act as derivatives,
\begin{equation}
p_A^n\Psi(z_n)=-\I\frac{\partial}{\partial z_n^A}\Psi(z_n^A),\qquad
\bar{p}_{A'}^n\Psi(z_n)=-\I\frac{\partial}{\partial\bar{z}^{A'}_n}\Psi(z_n).
\end{equation}
A generic state for all modes can be then pictured as a vector in an infinite tensor product, e.g.\ $\Psi(z_0,z_1,z_2,\dots)$. The obvious trouble with this construction is that the resulting Hilbert space $\mathcal{K}=\bigotimes_{n=0}^\infty\mathcal{H}_n$ is non-separable. This auxiliary (or kinematical) Hilbert space is therefore infinitely much larger than the ordinary Fock space, which is constructed over the lowest eigenstate of energy.  In our case, no preferred such vacuum state of quasi-local energy exists. A separable Hilbert space is found instead by considering the ground state of geometry (rather than quasi-local energy), and this \emph{\q{geometry vacuum}} will then select a separable subspace of the kinematical Hilbert space $\mathcal{K}$. 

Before going into the details of the construction, let us first better understand the structure of the Hamiltonian and its eigenvalues. Introducing the real and imaginary part of $L_0$, we have
\begin{equation}
H=L_0+\bar{L}_0,\qquad J=-\I(L_0-\bar{L}_0),\label{Hdef2}
\end{equation}
where the Hamiltonian $H$ generates dilatations in the radial direction $r=|z|$, while $J$ generates rotations around the origin $r=0$. In fact, the Hamiltonian $H$ defines a squeeze operator: going back to \eref{Lndef}, we have,
\begin{equation}
L_0=\frac{1}{4}\sum_{m\in\N_0}(2m+1)\big[p_A^mz^A_m+z^A_mp^m_A\big].\label{L0def}
\end{equation}
 The operator $z^A_mp_A^m=-\I z^A_m\partial/\partial z_A^m$ generates a dilation, and the eigenstates of $L_0$ can be constructed, therefore, from homogenous functions of the boundary spinors $z^A_m$. Such functions (if they are single valued) are labelled by two quantum numbers, namely a half integer $k_m\in\Z/2$ (the eigenvalue of $J$) and a real number $\rho_m\in\R$ (the eigenvalue of $H$). For a single mode, such homogenous wave functions satisfy
\begin{equation}
\forall\lambda\in\C-\{0\}:f^{(\rho,k)}(\lambda z^A)=\lambda^{\I\rho+k-1}\bar{\lambda}^{\I\rho-k-1}f^{(\rho,k)}(\lambda z^A).
\end{equation}

We thus see here quite explicitly that the quasi-local energy $H$ cannot be bounded from below, and we cannot use it to select a reasonable vacuum state. Yet, there is another and indeed more natural observable available at the boundary that we may use to select a sensible ground state, namely the length of a loop. The corresponding operator will select the ground state of geometry, which is the analogue of the Ashtekar\,--\,Lewandowski vacuum in the continuum. In our representation, this state can be constructed as follows. 

First of all, consider the circumference of an initial slice $\Sigma$ that cuts the cylinder $\mathcal{M}$ in half. The length of the boundary $\mathcal{C}=\partial\Sigma\simeq S^1$ with respect to the physical metric in the bulk is given by the integral,
\begin{equation}
\boldsymbol{L}[\mathcal{C}]=\oint_{\mathcal{C}}\di s\sqrt{g_{ab}(x(s))\dot{x}^a(s)\dot{x}^b(s)},
\end{equation}
where $\dot{x}^a(s)$ is the tangent vector to $\mathcal{C}$ as parametrised by $s\in S^1$. Now $\mathcal{C}$ lies inside the boundary $\mathcal{B}=\partial\mathcal{M}$, such that we can write the physical length in terms of the fiducial boundary metric $q_{ab}$ times the conformal factor $\Omega:\varphi^\ast_{\mathcal{B}}g_{ab}=\Omega^{-2}q_{ab}$. From the glueing conditions \eref{glucond} and \eref{Omdef} we then also know that the conformal factor is itself proportional to the norm squared of the boundary spinor, hence
\begin{equation}
\boldsymbol{L}[\mathcal{C}]=4\pi G\oint_{\mathcal{C}}\di s\sqrt{q_{ab}(x(s))\dot{x}^a(s)\dot{x}^b(s)}\,\|\xi(s)\|^2.
\end{equation}
Inserting the mode expansion \eref{xiseries}, we have
\begin{equation}
\boldsymbol{L}[\mathcal{C}]=4\pi G\sum_{m,n=-\infty}^\infty G_{AA'}^{mn}[\mathcal{C}]\xi^A_m\bar{\xi}^{A'}_n,\label{LHam2}
\end{equation}
where we introduced for any such loop $\mathcal{C}$ the following \emph{super-metric} on the covariant phase space,
\begin{equation}
G_{AA'}^{mn}[\mathcal{C}]:=\frac{1}{2\pi}\oint_{\mathcal{C}}\di s\Big|\frac{\di z(s)}{\di s}\Big|z^{m}(s)\bar{z}^n(s)\delta_{AA'}.\label{Smetrc}
\end{equation}
For any round circle $\mathcal{C}_R=\{z\in\C:|z|=R\}$ with respect to the fiducial boundary metric $q_{ab}$, this super-metric is diagonal,
\begin{equation}
G_{AA'}^{mn}[\mathcal{C}]=R^{2n+1}\delta^{mn}\delta_{AA'}.
\end{equation}
It is easy to check then that this hermitian form is  compatible with the symplectic structure (\hyperref[PoissStrct1]{48a--c}). The length of the loop $\mathcal{C}_R$ has turned therefore into an ordinary quadratic Hamiltonian on phase space, and the corresponding operator at the quantum level is diagonalised by introducing the following Landau operators,
\begin{subequations}\begin{align}
a^A_n[\mathcal{C}_R]&=\frac{1}{\sqrt{2}}\Big[R^{n+\frac{1}{2}}z^A_n+\frac{\I}{R^{n+\frac{1}{2}}}\delta^{AA'}\bar{p}_{A'}^{n}\Big],\\
b^A_n[\mathcal{C}_R]&=\frac{1}{\sqrt{2}}\Big[\frac{1}{R^{n+\frac{1}{2}}}p^A_{n}+\I R^{n+\frac{1}{2}}\delta^{AA'}\bar{z}_{A'}^{n}\Big]=:a^{A}_{-n-1},
\end{align}%
\label{landaus}%
\end{subequations}%
with $n=0,1,2,\dots$. 
It is immediate to check that these oscillators satisfy the canonical commutation relations,
\begin{equation}
\big[a^A_m,a_{Bn}^\dagger\big]=\big[b^A_m,b_{Bn}^\dagger\big]=\delta^{A}_B\delta_{mn},\label{hberg}
\end{equation}
and all other commutators among the oscillators vanish. Notice also that the hermitian conjugate is defined here both with respect to the Hilbert space inner product and the internal $SU(2)$ inner product, e.g.,
\begin{equation}
a^\dagger_{An}=\delta_{AB'}\bar{a}^{B'}_n.
\end{equation}
The length of the loop $\mathcal{C}$ is now nothing but the sum of all number operators. Choosing a normal ordering, we have
\begin{equation}
\normord{\boldsymbol{L}[\mathcal{C}_R]}\,=4\pi G\sum_{n=-\infty}^\infty a^\dagger_{An}[\mathcal{C}_R]a^A_n[\mathcal{C}_R].
\end{equation}
The vacuum state of geometry over a loop of radius $R$ is then simply given by the state that lies in the kernel of all annihilation operators,
\begin{equation}
a^A_n[\mathcal{C}_R]\big|0,{\mathcal{C}_R}\big\rangle=b^A_n[\mathcal{C}_R]\big|0,{\mathcal{C}_R}\big\rangle=0.\label{vacdef}
\end{equation}
This state is the boundary field theory analogue of the Ashtekar\,--\,Lewandowski vacuum, which is the state of vanishing geometry in the bulk \cite{Ashtekar:1993wf,Ashtekar:1994mh}. A complete and orthonormal basis in the resulting boundary Fock space $\mathcal{F}[{\mathcal{C}_R}]$ can be written in terms of spins $j_n\in\Z/2$ and magnetic indices $m_n=-j_n,\dots ,j_n$ for each mode, namely
\begin{align}\nonumber
\big|\{j_n,m_n\},\mathcal{C}_R\big\rangle=\prod_{n=-\infty}^\infty&\frac{1}{\sqrt{(j_n-m_n)!}}\frac{1}{\sqrt{(j_n+m_n)!}}\\
&\times
\big(a^\dagger_{0n}[\mathcal{C}_R]\big)^{j_n+m_n}\big(a^\dagger_{1n}[\mathcal{C}_R]\big)^{j_n-m_n}\big|0,{\mathcal{C}_R}\big\rangle
.\label{basisvec}
\end{align}
Each such state is an eigenvector of the length operator,
\begin{equation}
\normord{\boldsymbol{L}[\mathcal{C}_R]}\,\big|\{j_n,m_n\},\mathcal{C}_R\big\rangle=8\pi G\Big(\sum_{n=-\infty}^\infty j_n\Big)\big|\{j_n,m_n\},\mathcal{C}_R\big\rangle.
\end{equation}
Notice also that the spins $\{j_n\}_{n\in\Z}$ are all distinguishable, hence every such eigenvalue has infinite degeneracy.

The Fock space over the no-geometry vacuum state \eref{vacdef} does not quite yet define the physical Hilbert space. This is a consequence of the residual gauge transformations: rigid $SU(2)$ gauge transformations that are generated by constant gauge elements $\Lambda^i:\partial_a\Lambda^i=0$ preserve the gauge fixing condition \eref{gaugefix}. Physical states belong to the invariant subspace, which is obtained by a projector $P$,
\begin{equation}
\mathcal{F}_{\mathrm{phys}}[\mathcal{C}_R]=P\mathcal{F}[\mathcal{C}_R],\label{physHspace}
\end{equation}
which is defined by gauge averaging,
\begin{equation}
P:=\int_{SU(2)}d^3\mu_\Lambda\,\mathrm{exp}\big(-\I\Lambda^iS_{i}\big).\label{Pdf}
\end{equation}
In here, $d^3\mu$ is the Haar measure on $SU(2)$ and $S_{i}$ denotes the intrinsic spin of spacetime, namely the generator of rigid and \emph{internal} $SU(2)$ transformations,
\begin{equation}
S_{i}=\frac{1}{2}\sum_{n=-\infty}^\infty\ou{\sigma}{A}{Bi}a^\dagger_{An}a^B_n.\label{spindef}
\end{equation}
Notice in particular that the Fock vacuum \eref{vacdef} is itself an element of the physical Hilbert space. 

\subsection{Time evolution as squeezing, operator ordering}\label{sec3.3}
\noindent Let us now better understand the action of the Hamiltonian on the Hilbert space of the theory. If we go back to the definition of the quasi-local observables \eref{Lndef} we can express the Hamiltonian $H=L_0+\bar{L}_0$ in terms of creation and annihilation operators obtaining a two-mode squeeze operator,
\begin{equation}
H=\frac{1}{2}\sum_{n=0}^\infty(2n+1)\big(\epsilon_{AB}a^A_nb^B_n+\epsilon^{AB}a^\dagger_{An}b^\dagger_{Bn}\big).\label{hamdef}
\end{equation}
The (orbital) angular momentum, on the other hand, is the difference of two number operators
\begin{equation}
J=\frac{1}{2}\sum_{n=0}^\infty(2n+1)\big(a^\dagger_{An}a^A_n-b^\dagger_{An}b^A_n\big),
\end{equation}
where the oscillators $a^A_n$ and $b^A_n$ are defined as in \eref{landaus}. If we then act with the evolution operator on our reference vacuum state, we generate a highly entangled state,
\begin{align}\nonumber
&\E^{-\I\beta H}\big|0,\mathcal{C}_R\rangle=\\
&\; = \bigotimes_{n\in\N_0}\sum_{j_n\in\frac{1}{2}\N_0}\frac{\tanh^{2j_n}((n+\tfrac{1}{2})\beta)}{\cosh^2((n+\tfrac{1}{2})\beta)}\sum_{m_n=-j_n}^{j_n}\E^{-\I\pi m_n}\big|j_n,m_n\rangle_a\otimes\big|j_n,-m_n\rangle_b,\label{newvac1}
\end{align}
where the $SU(2)$ spin $j$ basis states $|j,m\rangle$ for the oscillators $a^A_n$ and $b^A_n$ are defined as in \eref{basisvec}. The Hamiltonian $H$ generates radial dilatations $z\rightarrow\E^{\beta} z$, and any radially shifted vacuum state \eref{newvac1} is then simply the geometric vacuum over the radially translated circle. In other words,
\begin{equation}
|0,\mathcal{C}_{R'}\rangle=\E^{-\I\beta H}\big|0,\mathcal{C}_R\rangle\quad\text{for}\quad R'=\E^{-\beta}R.\label{newvac2}
\end{equation}
It is easy to check that the radially translated vacuum state $|0,\mathcal{C}_{R'}\rangle$ has vanishing overlap with any vector in the original Fock space over $|0,\mathcal{C}_R\rangle$. In fact, the Hamiltonian does not preserve the Fock space, but maps it into $\mathcal{F}[\mathcal{C}_{R'}]=\exp(-\I\beta H)\mathcal{F}[\mathcal{C}_R]$, which is the Fock space over  the vacuum \eref{vacdef} for the radially translated loop $\mathcal{C}_{R'}$. For every different cross section $\mathcal{C}_R$ and $\mathcal{C}_{R'}$  of the boundary ($R\neq R'$), there is a different Fock vacuum. Each of these different Fock vacua represent different (and orthogonal) superselection sectors of  the kinematical and non-separable Hilbert space $\mathcal{K}=\bigotimes_{n=0}^\infty\mathcal{H}_n$. The radial \emph{\q{time evolution}} does not preserve any such Fock space, and the evolution operator $\mathrm{exp}(-\I\beta H)$ should be understood, therefore, on the larger (but non-separable) kinematical Hilbert space $\mathcal{K}$. The situation is in complete analogy with what happens for the Unruh effect: the vacuum state for a uniformly accelerated observer can be obtained from the Minkowski vacuum through a Bogoliubov transformation mixing the original creation and annihilation operators. The resulting Fulling\,--\,Rindler boosted vacuum is a highly entangled state that has vanishing overlap with the original Minkowski vacuum \cite{FullingQFT}. 

The usual CFT correlators can be now obtained through a limit of infinite squeezing. The \emph{\q{in}} and \emph{\q{out}} vacua in the infinite (euclidean) future and past are given by totally squeezed states,
\begin{equation}
\big|0,-\infty\rangle=\lim_{R\rightarrow 0}\big|0,\mathcal{C}_R\big\rangle,\qquad\big|0,+\infty\rangle=\lim_{R\rightarrow \infty}\big|0,\mathcal{C}_R\big\rangle.
\end{equation}
In the limit of infinite squeezing, the vacua for $R=0$ and $R=\infty$ turn into formal eigenstates of either position or momentum, namely
\begin{subalign}
p^A_n\big|0,-\infty\rangle=\bar{p}^{A'}_n|0,-\infty\rangle=0,\label{infsqueeze}\\
z^A_n\big|0,+\infty\rangle=\bar{z}^{A'}_n|0,+\infty\rangle=0.
\end{subalign}
The resulting $n$-point functions of the boundary CFT are given formally by
\begin{equation}
\big\langle \xi^A(z)\xi^B(w)\cdots\big\rangle=\frac{\big\langle 0,+\infty\big|R\big(\xi^A(z)\xi^B(w)\cdots\big)\big|0,-\infty\big\rangle}{\big\langle 0,+\infty\big|0,-\infty\big\rangle},
\end{equation}
where $R(\dots)$ denotes the radial ordering. The equations (\hyperref[infsqueeze]{80a,b}) for the \emph{\q{in}} and \emph{\q{out}} vacua,  determine all correlation functions, in particular,
\begin{equation}
\big\langle \xi^A(z)\xi^B(w)\big\rangle=\frac{1}{2\pi\I}\frac{\epsilon^{AB}}{z-w}.
\end{equation}

The choice of vacuum is now immediately related to the choice of operator ordering. Normal ordering amounts to subtracting otherwise divergent vacuum fluctuations from operator products. Since the radial evolution mixes creation and annihilation operators, normal ordering depends now on the chosen vacuum. In our case, we have chosen a reference state over $\mathcal{C}_R$, for some finite $R\neq 0$, and we thus remove the vacuum fluctuations of this particular reference state, namely
\begin{equation}
\normord{\xi^A(z)\bar{\xi}^{A'}(z)}\,=\xi^A(z)\bar{\xi}^{A'}(z)-\big\langle 0,\mathcal{C}_R\big|\xi^A(z)\bar{\xi}^{A'}(z)\big|0,\mathcal{C}_R\big\rangle.\label{norder}
\end{equation}

\subsection{Coherent states}
\noindent In section \eref{sec3.1} (see also equation \eref{basisvec}), we introduced a complete basis for the Fock space over a loop $\mathcal{C}_R$. Every element of this basis is an eigenstate of the length operator. There is minimal uncertainty in length, but the conjugate variable, which is an angle, is widely spread. Semi-classical states, on the other hand, should saturate the Heisenberg uncertainty relations, and cannot be, therefore, eigenstates of geometry. The simplest such states are given by the usual harmonic oscillator coherent states,
\begin{equation}
\big|\alpha,{\mathcal{C}_R}\big\rangle=\E^{-\frac{\|\alpha\|^2}{2}}\mathrm{exp}\Big(\sum_{n\in\Z}\alpha^{A}_na^\dagger_{An}\Big)\big|0,{\mathcal{C}_R}\big\rangle,\label{koharnt}
\end{equation}
where $\|\alpha\|^2$ denotes the $SU(2)$ norm of the boundary spinor $\{\alpha^A_n\}_{n\in\Z}$,
\begin{equation}
\|\alpha\|^2:=\sum_{n=-\infty}^\infty\alpha^{A}_n\bar{\alpha}^{A'}_n\delta_{AA'}.
\end{equation}

Using normal ordering \eref{norder}, the expectation values of the length operator for any cross section $\mathcal{C}$ of the boundary $\mathcal{B}$ reproduce the results from the classical theory,
\begin{equation}
4\pi G\oint_{\mathcal{C}}\left|\di z\right|\delta_{AA'}\big\langle\alpha,{\mathcal{C}_R}\big|\!\normord{\xi^A(z)\bar{\xi}^{A'}(z)}\!\big|\alpha,{\mathcal{C}_R}\big\rangle=4\pi G\oint_{\mathcal{C}}\big|\di z\big|\,\delta_{AA'}\alpha^A(z)\bar{\alpha}^{A'}(z),\label{expval}
\end{equation}
where $\alpha^A(z)$ is the expectation value of the field operator,
\begin{equation}
\alpha^A(z)=\langle\alpha,{\mathcal{C}_R}|\xi^A(z)|\alpha,{\mathcal{C}_R}\rangle.
\end{equation}
The coherent state \eref{koharnt} does not yet define an element of the physical Hilbert space \eref{physHspace}, because the state $|\alpha,\mathcal{C}_R\rangle$ is not invariant under the residual and global $SU(2)$ gauge transformations (see \hyperref[gaugefoot]{footnote 12}). The actual physical state is obtained as in \eref{Pdf} by projecting the state down onto its spin $j=0$ component, 
\begin{equation}
\big|\alpha,{\mathcal{C}_R}\big\rangle_{\text{phys}}=\int_{SU(2)}d^3\mu_\Lambda\,\mathrm{exp}\big(-\I\Lambda^iS_i\big)\big|\alpha,{\mathcal{C}_R}\big\rangle,\label{koharnt2}
\end{equation}
where $S_i$ is the generator of the residual gauge transformations \eref{spindef}. Since the length operator is invariant under such global frame rotations, the expectation values \eref{expval} for the length operator are unaffected by the gauge averaging. 

Finally, we may now construct a semi-classical state $|\alpha,\mathcal{C}_R\rangle$ that represents a catenoid. Having found an explicit solution \eref{xisolved} for the boundary spinor $\xi^A$ at the classical level, we can build a corresponding semi-classical state by simply exciting just the lowest modes,
\begin{equation}
\alpha^A_0=\pm\sqrt{\frac{\rho}{4 G}}\begin{pmatrix}0\\1\end{pmatrix},\quad\alpha^A_{-1}=\pm\sqrt{\frac{\rho}{4 G}}\begin{pmatrix}1\\0\end{pmatrix},
\end{equation}
while all other $\alpha^A_n$ vanish. These coherent states are a realisation in the continuum of the coherent states that were proposed at the discrete level by Livine and Speziale for quantum gravity in four dimensions, see \cite{Livine:2007vk}.

\subsection{Relation to loop quantum gravity}
\noindent Let us now explain the relation to loop quantum gravity \cite{ashtekar, rovelli, thiemann}. In loop quantum gravity, the quantum states for our cylindrical geometry are given by superposition of spin network states on an initial hypersuface $\Sigma$ of the cylinder $\mathcal{M}\simeq\Sigma\times\R$. Since we now have a boundary $\partial\Sigma\simeq S^1$, some of the edges of the spin network in the bulk will hit the boundary, where they excite a surface charge, namely an $SU(2)$ boundary spinor. On a fixed graph $\Gamma$,\footnote{A graph $\Gamma$ is an ordered list of oriented paths (edges $(e_1,e_2,\dots)=E_\Gamma$) on the disk $\Sigma$ that hit the boundary $\partial\Sigma\simeq S^1$ in a number of punctures $(p_1,p_2,\dots)$. We assume that every puncture belongs to only one such edge.} the state in the corresponding bulk plus boundary kinematical Hilbert space is given by an entangled state \cite{Donnelly:2016auv,Donnelly:2008vx}, 
\begin{equation}
\Psi\in\mathcal{H}^\Gamma_{\text{kin}}=\mathcal{H}_\Gamma\otimes_{SU(2)}\mathcal{H}_{\partial \Gamma},\label{kinspace}
\end{equation}
where the Hilbert space in the bulk
\begin{equation}
\mathcal{H}_\Gamma=L^2\big(SU(2)^{|E_\Gamma|}/_\Gamma \,SU(2)^{|V_\Gamma|}\big)
\end{equation}
is the space of square integrable functions on a number of copies of $SU(2)$ ($|E_\Gamma|$ denoting the number of edges) modulo gauge invariance at the vertices (there are $|V_\Gamma|$ of them). The boundary Hilbert space, on the other hand, is built from the excitations of the boundary charges that satisfy the commutation relations of the harmonic oscillator,\footnote{The simplest way to introduce these variables is to work in the spinorial representation of LQG, where every Wilson line is cut into two \emph{\q{half edges}}, each carrying a spinor that satisfies the commutation relations of the harmonic oscillator, see \cite{twist,komplexspinors,spinrep,Bianchi:2016tmw,Bianchi:2016hmk}.}
\begin{equation}
\big[a^A(p),a^\dagger_B(p')\big]=\delta_{pp'}\delta^A_B,\label{bcharg}
\end{equation}
where $\delta_{pp'}$ is the Kronecker delta for all punctures $p,p'\in\partial\Gamma$. A complete basis in the bulk plus boundary Hilbert space can be now written as follows,
\begin{align}\nonumber
\big|\Gamma,\{\iota_v\},\{j_e\}\big\rangle=\prod_{v\in V_\Gamma}{[\iota_v]}_{\vec{m}_v}\prod_{e\in E_\Gamma}&{\big[D^{(j_e)}(h^\dagger_e)\big]}^{m^+_em^-_e}\\
&\prod_{p\in \partial \Gamma}\frac{\big(a^\dagger_0(p)\big)^{j_p-m_p}}{\sqrt{(j_p-m_p)!}}\frac{\big(a^\dagger_1(p)\big)^{j_p-m_p}}{\sqrt{(j_p+m_p)!}}\big|0_\Gamma\big\rangle\otimes\big|0_{\partial\Gamma}\big\rangle,\label{state}
\end{align}
where $h^\dagger_e$ is a creation operator that excites a Wilson line for the spin connection in the bulk and all repeated bulk and boundary magnetic indices are summed over.\footnote{The multi-index $\vec{m}_v$ collects all magnetic indices $(m^{\pm}_{e},m^{\pm}_{e'},\dots)$ from incoming  (outgoing)  edges $(e, e',\dots)$ adjacent to a vertex $v\in V_\Gamma$, and $[D^{(j)}(U)]^{mn}$ is the spin $j$ Wigner matrix $D^{(j)}(U)^{mn}=(-1)^{j-n}\langle j,m|U|j,-n\rangle$. In addition, $(j_p,m_p)=(j_e,m_e^\pm)$ depending on wether the boundary edge $e$ intersects the puncture $p\in\partial\Gamma$ as an incoming $(m_e^+)$ or outgoing $(m_e^-)$ edge.} 
In the bulk, gauge invariance is imposed at the vertices: the coffeicients $[\iota_v]_{\vec{m}_v}$ (intertwiners) belong to the $SU(2)$ invariant subspace of all spin $j$ representations adjacent to the vertex. At the boundary, local $SU(2)$ gauge invariance follows from the contraction of the boundary spinors with the open legs of the spin networks ending at the boundary (hence the spin $j_p$ at every puncture matches the spin $j_e$ of the corresponding edge). This is the same kind of bulk to boundary coupling that can also be found on an isolated horizon, see for instance \cite{Ashtekar:2000eq, Bodendorfer:2013jba,Ghosh:2014rra}. 


The equations of motion in the bulk are the flatness constraint and the torsionless condition. At the quantum level, the torsionless condition translates into the requirement that the state is invariant under local bulk plus boundary $SU(2)$ gauge transformations. The flatness constraint, on the other hand, implies that the holonomy around any contractible loop $\alpha$ is given by the identity $h_\alpha[A]=\mathrm{Pexp}(-\oint_\alpha A)=\mathrm{id}$, and it is imposed at the quantum level simply by inserting a product of delta functions into the definition of the inner product \cite{3dimlqg,alexreview}, namely the kernel
\begin{equation}
K:=\prod_{c\in C_\Gamma}\sum_{j\in\N_0/2}(2j+1)\chi_j(h^\dagger_c)=\prod_{c\in C_\Gamma}\delta_{SU(2)}(h_c^\dagger).\label{kerneldef}
\end{equation}
The physical inner product is given then for any $\Psi,\Psi'\in\mathcal{H}_{\text{kin}}^\Gamma$ by
\begin{equation}
\langle\Psi|\Psi'\rangle_{\text{phys}}=\langle \Psi|K\Psi'\rangle.
\end{equation} 
In \eref{kerneldef}, the product goes over a set of fundamental cycles of the graph $\Gamma$ and $h_c$ is the path ordered product of all holonomies $h_{e},h_{e'},\dots$ around such a cycle $c$, e.g.\ for $c=e^{-1}\circ e'$, $h_c=h_{e'}h_e^{-1}$.  Taking also into account the gauge invariance of the entire bulk plus boundary state, we may then represent the elements of the resulting physical Hilbert space $\mathcal{H}^\Gamma_{\text{phys}}=K\mathcal{H}^\Gamma_{\text{kin}}$ as elements of the boundary Hilbert space alone,\footnote{This can be shown most easily as follows: Introduce a spanning tree $T$ of $\Gamma$ rooted at some fiducial puncture of the boundary ($T$ is a connected subgraph of $\Gamma$ that explores the entire graph: it contains no cycles but contains all punctures and every node of $\Gamma$). The links of $\Gamma$ that do not belong to $T$ (the \emph{leaves} of the tree) are in one-to-one correspondence with the elements of a set of fundamental cycles of $\Gamma$. Since we can never close a loop in a tree, we can then always find a gauge transformation that turns all holonomies along the links of $T$ (the \emph{branches} of the tree) to the identity. The flatness constraint sets then also the holonomies on the remaining links (i.e.\ the leaves of $T$) to the identity. See \cite{Dittrich:2014wda} for more details.}
\begin{align}
\big|(p_1,\dots,p_N),I\big\rangle= I^{m_1\dots m_N}\prod_{n=1}^N\frac{\big(a^\dagger_0(p_n)\big)^{j_n-m_n}}{\sqrt{(j_n-m_n)!}}\frac{\big(a^\dagger_1(p_n)\big)^{j_n-m_n}}{\sqrt{(j_n+m_n)!}}\big|0_{\partial\Gamma}\big\rangle\in\mathcal{H}_{\text{phys}}^\Gamma,\label{state2}
\end{align}
where we sum over all repeated magnetic indices and 
\begin{equation}
I\in \mathrm{Inv}_{SU(2)}\big(j_{1}\otimes\cdots\otimes j_{N}\big)
\end{equation}
is an element of the $SU(2)$ invariant subspace of the tensor product of the $SU(2)$ representations associated to each puncture. By imposing the flatness constraint in the bulk, the entire spin network has essentially collapsed into a single intertwiner $I$. 

We are now left to explain the relation to our construction in the continuum. Our strategy will be to start from the Fock space of the conformal boundary field theory and realise within this Fock space a representation of the boundary states \eref{state2}. 

First of all, we introduce a fiducial angular coordinate $\varphi$ at the boundary $\partial\Sigma=\mathcal{C}$ such that the $n$-th puncture\footnote{We may need to first reorder the enumeration of the punctures accordingly.} is located at the coordinate value
\begin{equation}
\varphi(p_n)=\varphi_n,\qquad 0\leq \varphi_1<\varphi_2<\dots<\varphi_N<2\pi.
\end{equation}
Consider then the following time-independent (Schrödinger) operator, 
\begin{equation}
\tilde{a}^A(\varphi)=\frac{1}{\sqrt{2\pi}}\sum_{n=0}^\infty\Big(a^A_n\E^{\I n\varphi}+b^A_n\E^{-\I(n+1)\varphi}\Big),\label{smeardop}
\end{equation}
which is constructed from the Landau operators $a^A_n$ and $b^A_n$ defined as in \eref{landaus}.  Next, we smear this operator over an interval containing the $n$-th puncture,\footnote{It is worth noting that in the $N\rightarrow\infty$ continuum limit, this definition returns Thiemann's regularisation of Dirac spinors \cite{Thiemannfermihiggs,ThiemannQSDV} in terms of half densities.} 
\begin{equation}
\tilde{a}^A(p_n)=\frac{1}{\sqrt{\phi_{n+1}-\phi_n}}\int_{I_n}\di\varphi\, a^A(\varphi),\label{smeard}
\end{equation}
where we divided the circle into intervals $I_n=[\phi_{n},\phi_{n+1})$, such that every puncture falls into only one such interval $\varphi_n\in(\phi_{n},\phi_{n+1})$. 

From the commutation relations \eref{hberg} of the Landau operators, we can then infer the commutation relations for the smeared oscillators \eref{smeardop}, namely 
\begin{equation}
\big[\tilde{a}^A(p_n),\tilde{a}^\dagger_B(p_{n'})\big]=\delta_{nn'}\delta^A_B.
\end{equation}
 
These are the same commutation relations that we had for the loop gravity boundary charges \eref{bcharg}, and we can therefore translate immediately the definition of the boundary states \eref{state2} on a graph $\Gamma$ into a definition for states in the physical Hilbert space of the boundary conformal field theory: we just have to replace in the definition of the loop gravity physical states \eref{state2} the creation operators ${a}_A^\dagger(p)$ for each puncture $p$ by the smeared operators $\tilde{a}_A^\dagger(p)$ that now excite the quanta of geometry out of the Fock vacuum \eref{vacdef} of the boundary conformal field theory. The physical Hilbert space of euclidean and three-dimensional quantum gravity on a graph $\Gamma$, can be realised therefore within the physical Hilbert space \eref{physHspace} of the conformal boundary field theory in the continuum. There is no continuum limit to be taken at this point, simply because we already have the physical Hilbert space in the continuum at hand (namely the $SU(2)$ invariant subspace of the Fock space \eref{physHspace} of the boundary conformal field theory), and we also know how to realise within this Fock space any state in the discrete graph Hilbert space $\mathcal{H}_{\text{phys}}^\Gamma$.\vspace{0.1em}

An altogether different question is how to compare our boundary CFT within the wider context of the spinfoam and GFT literature. The main difficulty with this question has to do with the conformal boundary conditions (\hyperref[bndryb]{2a,b}), which are not immediate to translate into the discrete spinfoam formalism. For example, Dittrich, Livine and collaborators have studied very recently \cite{Dittrich:2017hnl,Dittrich:2017rvb} the discrete boundary theory that is generated by the spin network evaluation of the Ponzano\,--\,Regge amplitudes for Dirichlet boundary conditions (the intrinsic two-dimensional geometry is kept fixed at the boundary). At the present stage, it is difficult to compare our results with these developments, because the conformal boundary conditions  (\hyperref[bndryb]{2a,b}) depend on a fiducial background structure (namely the boundary metric $q_{ab}$), and it is not immediate to see how such a background structure should be encoded into specific spin network boundary states. Indeed, the usual Dirichlet boundary conditions are comparably easy to impose within the spinfoam approach, but very little is known about more general boundary conditions at the discretised level, see \cite{Dittrich:2017hnl,Dittrich:2017rvb} and \cite{Oriti:2002hv,OLoughlin:2000yww,Arcioni:2001ds}. For conformal boundary conditions (on an e.g.\ square lattice), it would be necessary to fix the angles between adjacent boundary edges. Such a condition is difficult to impose in the spin network representation, because the angles turn into quantum operators that do not commute among themselves. The best one could do is to choose specific coherent boundary states, such that the angles between neighbouring edges are peaked at values that would correspond to some given boundary metric $q_{ab}$. A further difficulty is that the conformal boundary conditions (\hyperref[bndryb]{2a,b}) not only fix the conformal class of the boundary metric, but also require that the trace of the extrinsic curvature vanishes. This second boundary condition \eref{bndrya} is equivalent to the holomorphicity of the boundary spinors, see \eref{EOMb}. To impose such a holomorphicity condition at the discrete level of a spin network graph, we would have to first discretise the  Cauchy\,--\,Riemann differential equations for the boundary spinors and replace them by appropriate difference equations on the square lattice. Such a program would be worthwhile to initiate, because our results also suggest possible relations within the context of group field theory (GFT), see e.g.\ \cite{Oriti:2014aa}. In fact, very recently a quantum cosmological GFT model \cite{GFTQC} has been introduced, where the cosmological evolution is generated by a squeezing Hamiltonian that is very similar structurally to the Hamiltonian \eref{hamdef} of the boundary CFT. It would be important to investigate if there are more profound such connections, which could be helpful to characterise the GFT amplitudes (and their critical points) in terms of conformal boundary field theories.


\subsection{Evaluation of the quasi-local partition function}
\noindent Finally, we want to discuss some strategies to calculate the generalised euclidean partition function. The main purpose of the section is to show that we can make sense of the following Virasoro character\footnote{The usual partition function $\mathrm{Tr}_{\mathrm{phys}}(\E^{-\beta H-\varphi J})$ is obtained then by analytic continuation.}
\begin{equation}
Z(\beta,\varphi)=\operatorname{Tr}_{\text{phys}}\big(\E^{-\I\beta H-\I\varphi J}\big)=\int_{SU(2)}d^3\mu_\theta\operatorname{Tr}_{\mathcal{F}}\big(\E^{-\I\beta H-\I\varphi J}\E^{-\I\theta^iS_i}\big),\label{Zustandssumme}
\end{equation}
despite the fact that the quasi-local Hamiltonian is not positive definite. Notice that the trace of the evolution operator $U(\beta,\varphi)=\E^{-\I\beta H-\I\varphi J}$ is taken here in the \emph{physical} Hilbert space \eref{physHspace}, which is the subspace of vanishing spin in the Fock space $\mathcal{F}$ over the Ashtekar\,--\,Lewandowski  vacuum \eref{vacdef}. As it stands, the expression \eref{Zustandssumme} for the state sum is badly singular: it diverges for $\beta=0$, and vanishes for all other values of $\beta$. To regularise it, we introduce a sharp UV cutoff, and take only those oscillators $a^A_n$ and $b^A_n$ into account that lie below the cutoff, 
\begin{equation}
Z_N(\beta,\varphi,{\theta})=\prod_{n=0}^N\operatorname{Tr}_{\mathcal{H}_n}\big(\E^{-\I\beta H^{(n)}-\I\varphi J^{(n)}}\E^{-\I\theta^iS_i^{(n)}}\big),\label{Zustandssumme2}
\end{equation}
where the trace is now taken for the $n$-th mode in the Hilbert space $\mathcal{H}_n=L^2(\C^2,d^4z_n)$ and $\theta=|\vec{\theta}|$ is a class angle. To calculate the partition function \eref{Zustandssumme} we have to take the limit $N\rightarrow\infty$ and integrate over the residual gauge orbit of rigid and internal $SU(2)$ frame rotations at the boundary,
\begin{equation}
Z_N(\beta,\varphi)=\frac{1}{\pi}\int_0^{2\pi}\di\theta\,\sin^2\Big(\frac{\theta}{2}\Big)Z_N(\beta,\varphi,{\theta}).
\end{equation}
The regularised partition function can be now readily evaluated: consider the Hamiltonian $H^{(n)}$, and the orbital $J^{(n)}$  and spin angular momentum $S^{(n)}$ for each mode,
\begin{equation}
H^{(n)}=L_0^{(n)}+\bar{L}_{0}^{(n)},\quad J^{(n)}=-\I\big(L_0^{(n)}+\bar{L}_{0}^{(n)}\big),\quad S^{(n)}_i=\Sigma_i^{(n)}+\bar{\Sigma}_i^{(n)},
\end{equation}
where $L^{(n)}_0$ and $\bar{L}_0^{(n)}$ denote the complexified squeeze operators,
\begin{subalign}   
L^{(n)}_0&=-\frac{\I}{2}(2n+1)\Big(1+z^A_n\frac{\partial}{\partial z^A_n}\Big),\label{L0na}\\
\bar{L}^{(n)}_0&=-\frac{\I}{2}(2n+1)\Big(1+\bar{z}^{A'}_n\frac{\partial}{\partial\bar{z}^{A'}_n}\Big)\label{L0nb}.
\end{subalign}
The intrinsic spins, on the other hand, are given by
\begin{subalign} 
\Sigma^{(n)}_i&=-\frac{1}{2}\ou{\sigma}{A}{Bi}z^B_n\frac{\partial}{\partial z^A_n},\label{S0na}\\
\bar{\Sigma}^{(n)}_i&=+\frac{1}{2}\ou{\bar\sigma}{A'}{B'i}\bar{z}^{B'}_n\frac{\partial}{\partial \bar{z}^{A'}_n}\label{S0nb},
\end{subalign}
where $\ou{\sigma}{A}{Bi}$ are again the Pauli matrices.

So far, we have introduced two different bases in our boundary Hilbert space: in \eref{koharnt} we introduced a basis of coherent states, and the eigenbasis of the length operator was introduced in \eref{basisvec}. To compute the euclidean state sum \eref{Zustandssumme2} it is useful to introduce yet another (distributional) basis. Consider plane waves,
\begin{equation}
\langle z^A\big|k_A\rangle=\frac{1}{(2\pi)^2}\E^{-\I k_Az^A-\CC},
\end{equation}
which are normalised as 
\begin{equation}
\langle k_A\big|k_A'\rangle=\delta_{\C^2}(k_A-k_A'),
\end{equation}
with respect to the integration measure on $\C^2$,
\begin{equation}
d^4k:=\frac{1}{4}\di k_A\wedge\di k^A\wedge\di \bar{k}_{A'}\wedge\di \bar{k}^{A'}.
\end{equation}
The generators \hyperref[L0na]{(89a,b)} act as complexified dilations sending $z^A_n$ into $\E^{-(n+\frac{1}{2})(\beta-\I\varphi)}z^A_n$ (in addition they also generate an overall rescaling of the state) while the spin operators $\Sigma_i^{(n)}$ generate internal $SU(2)$ frame rotations. The relevant matrix elements are,
\begin{subalign}
\langle z^A|\E^{-\I \beta H^{(n)}-\I\varphi J^{(n)}}|k_A\rangle&=\E^{-(2n+1)\beta}\langle z^A|\E^{-(n+\frac{1}{2})(\beta-\I\varphi)}k_B\rangle,\\
\langle z^A|\E^{-\I \Lambda^iS_i^{(n)}}|k_A\rangle&=\langle z^A|k_B\ou{U}{B}{A}\rangle,
\end{subalign}
for $\ou{U}{A}{B}=\ou{[\exp(\frac{\I}{2}\Lambda^i\sigma_i)]}{A}{B}\in SU(2)$. For any $\beta\neq 0$ we now have,
\begin{align}\nonumber
\operatorname{Tr}_{\mathcal{H}_n}&\big(\E^{-\I\beta H^{(n)}-\I\varphi J^{(n)}}\E^{-\I\Lambda^iS_i^{(n)}}\big)=
\int_{\C^2}\! d^4k\,\big\langle k_A\big|\E^{-\I(\beta-\I\varphi)L_0^{(n)}}\E^{-\I(\beta+\I\varphi)\bar{L}_0^{(n)}}\E^{-\I\Lambda^iS_i^{(n)}}\big|k_A\big\rangle=\\\nonumber
&\quad=\E^{-(n+\frac{1}{2})(\beta-\I\varphi)}\E^{-(n+\frac{1}{2})(\beta+\I\varphi)}\int_{\C^2} d^4k\,\big\langle k_A\big|\E^{-(n+\frac{1}{2})(\beta-\I\varphi)}\ou{U}{B}{A}k_B\big\rangle=\\
\nonumber&\quad=\E^{-(2n+1)\beta}\int_{\C^2}\! d^4k\,\delta_{\C^2}\Big(\big(\delta^B_A-\E^{-(n+\frac{1}{2})(\beta-\I\varphi)}\ou{U}{B}{A}\big)k_B\Big)=\\
&\quad=\frac{\E^{-(2n+1)\beta}}{\left|\mathrm{det}\Big(\bbvar{1}-\E^{-(n+\frac{1}{2})(\beta-\I\varphi)}U\Big)\right|^2}.
\end{align}
The determinant in the denominator can be written as
\begin{equation}
\mathrm{det}\Big(\bbvar{1}-\E^{-(n+\frac{1}{2})(\beta-\I\varphi)}U\Big)=\big(1-q^{n+1/2}\E^{\I\theta/2}\big)\big(1-q^{n+1/2}\E^{-\I\theta/2}\big)
\end{equation}
for
\begin{equation}
q=\E^{-\beta+\I\varphi}\quad\text{and}\quad\mathrm{Tr}(U)=2\cos\frac{\theta}{2}.
\end{equation}
The regularised partition function \eref{Zustandssumme2} turns then into the product
\begin{equation}
Z_N(\beta,\varphi,\theta)=\prod_{n=0}^N\frac{\E^{-(2n+1)\beta}}{\left|\Big(1-q^{n+1/2}\E^{+\I\theta/2}\big)\big(1-q^{n+1/2}\E^{-\I\theta/2}\Big)\right|^2}.\label{Zustandssumme3}
\end{equation}
Assuming $\beta>0$, the denominator has a non-vanishing limit for $N\rightarrow \infty$. The troublesome contribution comes only from the nominator, which vanishes for $N\rightarrow\infty$ and $\beta>0$.  But this divergence is not unexpected (it is a zero-point vacuum contribution) and it must be regularised with other methods.\footnote{$\zeta$-function regularisation yields $[\prod_{n=0}^\infty\E^{-2(n+1/2)\beta}]_{\text{reg}}=\E^{-\frac{1}{12}\beta}$}

It seems intriguing that the whole expression looks somewhat similar to what was found by Witten and Maloney in an altogether different context \cite{Maloney:2007ud}. But there are also most crucial differences: first of all, we have an additional Lagrange multiplier $\theta$ in our partition function, which is dual to the intrinsic spin of spacetime  (we integrate over this Lagrange multiplier to impose the vanishing of torsion in the bulk). Such an angle is necessarily absent in any approach based on the metric formulation, where the torsionless condition is solved prior to quantisation. In addition, and this is more significant, Witten and Maloney's results for the partition function were derived from the asymptotic symmetries \cite{Brown:1986nw} of three-dimensional euclidean quantum gravity with negative cosmological constant. In our case, the cosmological constant is set to zero from the onset and the entire calculation happens in a quasi-local context, where the gravitational field is quantised in a finite region. The question is then not so much why we failed to reproduce the results of Witten and Maloney exactly, but to explain the rather curious similarities between the two approaches (given that the cosmological constant vanishes in our setting and the entire calculation happens at a finite boundary). The answer could lie in the peculiar simplicity of three-dimensional gravity: in three dimensions, there are no gravitational waves, and the only degrees of freedom of our bulk plus boundary system are given by (i) the shape and geometry of the boundary (encoded in the edge modes of the boundary spinor $\xi^A$), and (ii) global degrees of freedom that parametrise the moduli space of flat connections.\footnote{In our case, we are working on an infinite cylinder of fixed topology $\mathcal{M}\simeq\Sigma\times \R$, where $\Sigma$ is a disk in $\R^2$. The moduli space of flat connections on $\Sigma$ contains therefore just the trivial element.} In addition, the on-shell action for our bulk plus boundary system vanishes: the boundary term vanishes because the boundary is a minimal surface, and the bulk term vanishes because the curvature tensor vanishes for $\Lambda=0$. The vanishing of the on-shell bulk plus boundary action is significant, because it may suggest, in fact, that it does not cost any phase space volume to move the boundary outwards (towards infinity), and this may give an intuitive explanation for why we found such a similar result at the quasi-local level. 

\section{Conclusion and discussion}
\noindent Let us summarise and discuss the results of the paper. The first half of the paper was concerned with the classical bulk plus boundary field theory. In the interior of $\mathcal{M}$, the theory is defined by the usual triadic Palatini action for euclidean general relativity with vanishing cosmological constant. Working on an infinitely tall cylinder of finite circumference, we then had to introduce an appropriate boundary term for our boundary conditions (\hyperref[bndrya]{2a,b}). We then introduced such a boundary term in terms of an $SU(2)$ boundary spinor $\xi^A$, which is minimally coupled to the pull-back of the spin connection to the boundary. This spinor is not to be confused with a material field of reference. It is a purely geometric quantity that encodes the extrinsic and intrinsic geometry of the boundary. The squared $SU(2)$ norm $\|\xi\|^2=\langle\xi|\xi\rangle$, for example, determines the conformal factor $\Omega$ that relates the fiducial two-dimensional metric $q_{ab}$ at the boundary to the pull-back of the physical metric $g_{ab}$ in the bulk, the internal direction of the spinor, on the other hand, defines the tangent plane of the boundary at every point: the expectation value $\langle\xi|\vec{\sigma}|\xi\rangle/\|\xi\|^2$ gives a three-vector $\vec{n}$, which is the normal vector of the boundary with respect to the triadic basis, see \eref{ndef}.

Taking also into account the variations of the spinor at the boundary, we then found additional boundary equations of motion that define a conformal field theory with vanishing central charge.\footnote{This boundary field theory resembles the bosonic $\beta\gamma$-CFT for the ghosts in superstring theory.} Solutions to the boundary equations of motion are given by holomorphic spinors on the boundary. The holomorphicity of $\xi^A$ has a neat geometric interpretation, for it implies that the boundary is a minimal surface, which is just our boundary condition \eref{bndrya}. A minimal example for a non-trivial solution of the entire bulk plus boundary system was given in \hyperref[sec2.3]{section 2.3} by a holomorphic spinor \eref{xisolved} defined over a catenoid flatly embedded in $\R^3$. The next step ahead was to study the classical phase space and the gauge symmetries of the bulk plus boundary theory. 

At the Hamiltonian level, all bulk diffeomorphisms that vanish at the boundary (in addition to internal $SU(2)$ gauge transformations of the bulk plus boundary fields), are gauge symmetries (degenerate directions of the pre-symplectic two-form). On the other hand, active diffeomorphisms, whose pull-back to the boundary induce conformal maps of the fiducial boundary metric define actual boundary symmetries (motions) generated by the quasi-local charges \eref{Hdef3}, which satisfy the classical Virasoro algebra with vanishing central charge (this is also true at the quantum level provided we use the symmetric ordering for $H$ and $J$, as in \eref{L0def}). To quantise the classical phase space we then had to find a representation for the classical commutation relations at the quantum level. The difficulty was that there is no obvious choice of vacuum: choosing a vacuum amounts to choosing a complex structure on phase space, and in our case there is no natural such structure available. In quantum field theory, it is usually the (free) Hamiltonian that determines the complex structure and provides a natural definition for the vacuum of the (free) theory. In our case, the Hamiltonian $H=L_0+\bar{L}_0$ is a two mode squeeze operator, its spectrum is continuous and unbounded from below. There is therefore no obvious way to construct the Hilbert space from the excitations of the quasi-local energy. A possible solution to this trouble was found by looking at a different observable, namely the physical length of a circumference of the boundary. The length of a curve $\mathcal{C}\subset\mathcal{B}$ in the boundary is determined by the line integral of the conformal factor along $\mathcal{C}$. Since the conformal factor is given by the square of the boundary spinor (times the Planck length), this integral defines a hermitian form on the classical phase space. For a circular path (with respect to the fiducial background metric) this hermitian form can be diagonalised trivially by introducing an infinite tower of harmonic oscillators $a^A_n$ and $b^A_n$. The vacuum state for these oscillators describes a configuration where the boundary collapses into a point. Such a state is very well known from loop quantum gravity, where the Ashtekar\,--\,Lewandowski vacuum defines a totally degenerate geometry \cite{Ashtekar:1993wf,Ashtekar:1994mh}.  In the remaining part of the paper we studied some additional aspects of the resulting quantum theory: first of all we introduced coherent states that approximate the geometry of a classical solution of our boundary equations of motion (such as a catenoid, see \hyperref[sec2.3]{section 2.3}). Finally, we studied the euclidean partition function \eref{Zustandssumme}. The expression \eref{Zustandssumme} is singular, after some formal manipulations we arrived at a regularised expression. The result resembles the BMS characters that were studied for euclidean gravity in three dimensions, see \cite{Barnich:2015mui, Oblak:2015sea}, but there are also major differences, most notably the additional gauge average over the angle $\theta$, which is dual to the intrinsic spin of spacetime (such an angle $\theta$ is necessarily absent from any approach to quantum gravity based on the metric variables, where the torsionless condition is always solved prior to quantisation).

Let us also stress that the conformal field theory that we found at the boundary of our cylinder has a couple of rather unusual features: the fundamental configuration variable is a two-component spinor that satisfies bosonic commutation relations. The Virasoro algebra has a vanishing central charge, the quasi-local energy $H=L_0+\bar{L}_0$ is a two-mode squeeze operator (rather than a sum of number operators), and its spectrum is continuous and unbounded from below. In addition, and this is probably the most curious novelty of this paper, the conformal factor $\Omega$ turns into a number operator at the quantum level. This is in stark contrast to all perturbative approaches to quantum gravity, where the conformal factor $\Omega$ is always treated as a classical field (such as a radial coordinate $\Omega=O(r^{-1})$), which is sent to zero prior to quantisation. \vspace{0.7em}

\emph{Acknowledgments.} In general, I owe much to the discussions with Eugenio Bianchi, Henrique Gomes and Rob Myers, and I would like to thank them for most usual comments and critical remarks. I have presented the results of this paper in February 2018 at the sixth quantum gravity workshop in Austria and I would like to thank the organisers and participants of this workshop for most useful feedback and discussions. I would like to thank in particular Simone Speziale and Norbert Bodendorfer. This research was supported from Perimeter Institute for Theoretical Physics. Research at Perimeter Institute is supported by the Government of Canada through the Department of Innovation, Science and Economic Development and by the Province of Ontario through the Ministry of Research and Innovation.

\providecommand{\href}[2]{#2}\begingroup\raggedright\endgroup

\end{document}